\documentclass[twocolumn,showpacs,prb,superscriptaddress]{revtex4}

\usepackage{psfrag}
\usepackage{enumerate}
\usepackage{graphicx}
\usepackage{amssymb}
\usepackage[tbtags]{amsmath}
\newcommand{\FeMn}{$\mathrm{Fe_{0.5}Mn_{0.5}TiO_3}$ }
\newcommand{\FeZnF}{$\mathrm{Fe_{0.85}Zn_{0.15}F_2}$ }

\begin{document}

\title{Finite versus zero-temperature hysteretic behavior of spin glasses:\\
Experiment and theory}

\author{Helmut G.~Katzgraber}

\affiliation
{Theoretische Physik, ETH Z\"urich,
CH-8093 Z\"urich, Switzerland}

\author{Didier H\'{e}risson}
\author{Michael \"{O}sth}
\author{Per Nordblad}

\affiliation
{Department of Engineering Sciences, Uppsala University,
Box 534, SE-751 21 Uppsala, Sweden}

\author{Atsuko Ito}
\author{Hiroko Aruga Katori}

\affiliation
{Riken, Hirosawa 2-1, Wako, Saitama, 351-0198, Japan}

\date{\today}

\begin{abstract}

We present experimental results attempting to fingerprint
nonanalyticities in the magnetization curves of spin glasses found by
Katzgraber {\em et al.}~[Phys.~Rev.~Lett.~{\bf 89}, 257202 (2002)] via
zero-temperature Monte Carlo simulations of the Edwards-Anderson Ising
spin glass. Our results show that the singularities at zero temperature
due to the reversal-field memory effect are washed out by the finite
temperatures of the experiments. The data are analyzed via the first
order reversal curve (FORC) magnetic fingerprinting method. The
experimental results are supported by Monte Carlo simulations of the
Edwards-Anderson Ising spin glass at finite temperatures which agree
qualitatively very well with the experimental results.  This suggests
that the hysteretic behavior of real Ising spin-glass materials is
well described by the Edwards-Anderson Ising spin glass. Furthermore,
reversal-field memory is a purely zero-temperature effect.

\end{abstract}

\maketitle

\section{Introduction}
\label{sec:introduction}

While the nonequilibrium behavior of spin
glasses\cite{binder:86,young:98,kawashima:03} has been
studied in detail and several aspects have been applied
to fields as wide as biology and financial analysis, less
work\cite{zhu:90,sethna:93,lyuksyutov:99,pazmandi:99,katzgraber:02b}
has been done to understand the hysteretic behavior\cite{bertotti:98}
of spin glasses in a time-dependent field. Because the understanding
of hysteretic systems plays a prominent role in the development
of magnetic recording media, understanding fundamental properties
of simple hysteretic systems such as spin glasses is of paramount
importance.

In Ref.~\onlinecite{katzgraber:02b} Katzgraber {\em et al.}~reported
on a novel memory effect---the reversal-field memory effect---found
in spin-glass systems: When the field is decreased from saturation
to a reversal field $H_{\rm R}$, upon return to saturation the
magnetization exhibits a singularity (kink) at $H = - H_{\rm R}$.
The underlying spin-reversal symmetry of the Hamiltonian is the
source of this effect which was observed first numerically at
zero temperature for the two-dimensional Edwards-Anderson Ising
spin glass.\cite{edwards:75} In contrast, systems which do not
possess spin-reversal symmetry, such as the random-field Ising
model,\cite{lyuksyutov:99,sethna:93} do not exhibit this memory
effect. Because the kinks in the magnetization curves are easily
overlooked, Katzgraber {\em et al.}~used the first order reversal
curve (FORC) method\cite{dellatorre:99,pike:99,katzgraber:02b} to
characterize the reversal-field memory effect.  In a FORC diagram,
the kinks in the magnetization curves are captured by a pronounced
vertical ridge along the coercivity axis (see below).

Using the superb fingerprinting abilities of the FORC method we
have attempted to experimentally detect the reversal-field memory
effect in \FeMn Ising spin-glass samples.  Our finite-temperature
experimental results show that the reversal-field memory effect is not
present. Moreover, these results are verified via finite-temperature
nonequilibrium Monte Carlo simulations of the three-dimensional
Edwards-Anderson Ising spin glass. The numerical results show that the
reversal-field memory effect can only be observed at experimentally
inaccessible temperatures.  The good agreement between the simulation
and experiments suggest that the hysteretic properties of \FeMn can
be well modeled using the short-range Edwards-Anderson Ising spin
glass model.

The paper is structured as follows. First we review the FORC method
in Sec.~\ref{sec:forc}. In Sec.~\ref{sec:experiments} we present
experimental results on $\mathrm{Fe_{0.5}Mn_{0.5}TiO_3}$. In Sec.~\ref{sec:numerics} results
of zero- and finite-temperature Monte Carlo simulations of the
three-dimensional Edwards-Anderson Ising spin glass are presented,
followed by concluding remarks in Sec.~\ref{sec:conclusions}.

\section{Outline of the FORC Method}
\label{sec:forc}

FORC diagrams,\cite{dellatorre:99,pike:99,katzgraber:02b} which
can be viewed as a model-independent generalization of Preisach
diagrams\cite{preisach:35,mayergoyz:86} complement current
methods to characterize magnetic interactions in hysteretic
systems,\cite{he:92,proksh:94,hedja:94} such as the $\delta M$
method.\cite{che:92,el-hilo:92} The advantage of the FORC method over
other approaches is the extreme sensitivity to microscopic details of
hysteretic systems. Although the wealth of information a FORC diagram
delivers still remains to be fully characterized, the method can be
used as an extremely sensitive ``fingerprint'' of a hysteretic system.

To calculate a FORC diagram, a family of FORCs with different
reversal fields $H_{\rm R}$ is measured either experimentally
or numerically.\cite{comment:forc} The measurements start at
saturation and the field is then reversed at different values
of the reversal field $H_{\rm R}$.  The mixed second-order
derivative\cite{dellatorre:99,pike:99} of the magnetization $M(H,
H_{\rm R})$ as a function of the applied and reversal field yields
the FORC diagram $\rho(H, H_{\rm  R})$ given by
\begin{equation}
\rho(H, H_{\rm  R})= -\frac{1}{2}
[{\partial}^2 M/{\partial} H {\partial} H_{\rm  R}] \, .
\label{eq:rho}
\end{equation}
In general, a rotation of the coordinates to
\begin{eqnarray}
H_{\rm c} &=& [H-H_{\rm R}]/2 \, ,\\
H_{\rm b} &=& [H+H_{\rm R}]/2 \, ,
\end{eqnarray}
the local coercivity $H_{\rm c}$ and bias $H_{\rm b}$, respectively,
yields the FORC distribution $\rho(H_{\rm b}, H_{\rm c})$. Therefore,
under the assumption that a hysteretic system can be described by
microscopic switching units, each of which have a given coercivity
$H_{\rm c}$ and a bias field $H_{\rm b}$, the FORC diagram
corresponds to the distribution of these microscopic bias/coercivity
fields. In the case of discrete spin models such as the random-field
Ising model,\cite{lyuksyutov:99,sethna:93} these microscopic
switching units can, in general, be identified with the simulated
spins.\cite{katzgraber:04e} This is not the case for systems which
exhibit frustration,  such as the Edwards-Anderson Ising spin-glass
model.  FORC diagrams have been applied to a variety of hysteretic
systems and fields ranging from geological applications\cite{pike:99}
to magnetic materials,\cite{katzgraber:02b,liu:05,davies:05} as well
as more exotic systems such as cyclic voltammetry,\cite{hamad:06}
superconducting perovskite materials,\cite{davies:05a} and thermal
hysteresis.\cite{enachescu:04}

\section{Experimental Results}
\label{sec:experiments}

\begin{figure}
\includegraphics [width=8cm] {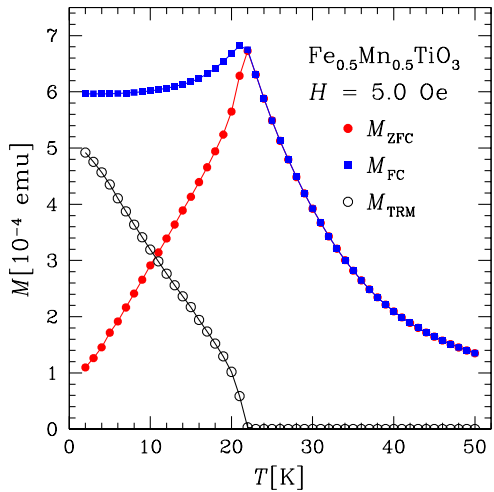}

\vspace*{-1.0cm}

\caption{(Color online) 
Zero field cooled (ZFC), field cooled (FC) and thermoremanent
magnetization (TRM) of the \FeMn sample as a function of temperature $T$
for an applied field of $H = 5$Oe. For this experimental sample the glass
transition occurs at $T_{\rm c} \approx 21$ K.}
\label{fig:MT}
\end{figure}

\begin{figure}
\includegraphics [width=8.0cm] {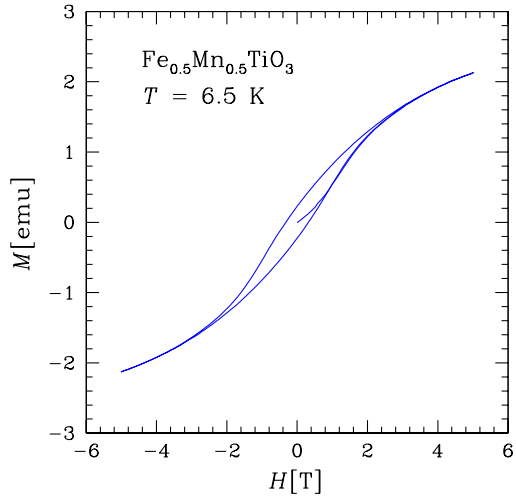}

\vspace*{-1.0cm}

\caption{(Color online)
Hysteresis curve of \FeMn at $6.5$ K, which corresponds to a
temperature of $T \approx 0.31 T_{\rm c}$ (see Fig.~\ref{fig:MT}).}
\label{fig:MH}
\end{figure}

The experiments have been performed on a single crystal of the Ising
spin glass \FeMn in a Quantum Design MPMS-XL 5T SQUID (superconducting
quantum interference device) magnetometer. Low-field magnetization vs
temperature curves in zero field cooled (ZFC), field cooled (FC),
and thermoremanent magnetization (TRM) protocols for \FeMn are
shown in Fig.~\ref{fig:MT}. The spin-glass transition temperature
of \FeMn is $T_{\rm c} \approx 21$K. These measurements are made
with the applied field parallel to the $c$ axis of the sample---the
corresponding curves measured perpendicular to the field show only a
weakly temperature-dependent paramagnetic response, no irreversibility
between the ZFC and FC magnetization and zero remanence.\cite{Ito:90}
In addition, in Fig.~\ref{fig:MH} a magnetization ($M$) vs applied
field ($H$) hysteresis curve measured at $6.5$ K, the temperature of
the FORC experiment described below, is shown.  An extensive study
of the hysteretic behavior of \FeMn has been reported by Ito {\em et
al.}~in Ref.~\onlinecite{ito:97}.

The FORC measurements have been made at as low a temperature as
possible to minimize influences on the FORC distribution from thermal
relaxation effects. The temperature must, however, be high enough for
the sample to have a reversible magnetization at the highest magnetic
field available ($5$ T). The chosen working temperature of $6.5$ K
($T/T_{\rm c} \approx 0.31$) optimally fulfills these two requirements.

\begin{figure}[!tbp]
\includegraphics[width=\columnwidth]{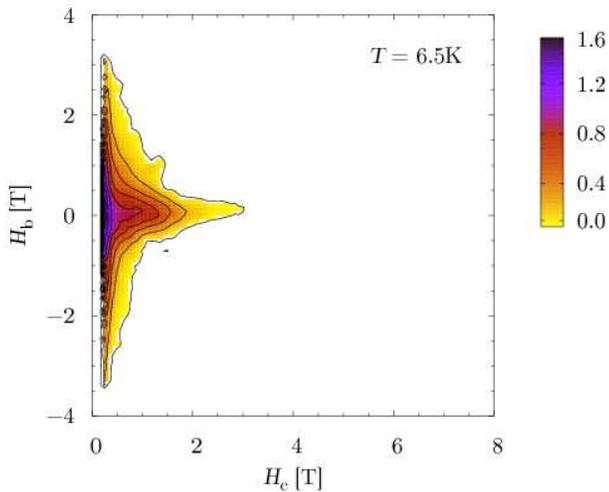}

\vspace*{-0.2cm}

\caption{(Color online)
FORC diagram of \FeMn at $T = 6.5$ K ($T/T_{\rm c} \approx 0.31$).
Note that the distribution is not normalized.
}
\label{fig:forc-exp}
\end{figure}

The derived FORC distribution is shown in Fig.~\ref{fig:forc-exp}. The
individual FORCs behind the distribution have been measured using field
steps of $80$ mT and a limiting field of $\pm 5$ T. The measurements
are made at a constant field increase rate governed by the time it
takes to change the field and record one data point (each data point
must be recorded at constant field in the SQUID).  To avoid some of
the influence of relaxation of the magnetization after decreasing
the field from the initial high value to $H_{\rm R}$, the sample is
kept for $1000$ s at the reversal field before the actual recording
of data during field increase starts. A preliminary report on these
results is found in Ref.~\onlinecite{herisson:07}. It can be noted
from Fig.~\ref{fig:forc-exp} that there is a wide distribution of
both coercivity and bias fields. The distribution shows a smooth
variation with a weak maximum at a coercivity of about $1$ T and a bias
field close to zero; and it only slowly decays to zero from a broad
ridge along the coercivity axis. In particular, the horizontal ridge
reminiscent of the reversal-field memory effect\cite{katzgraber:02b}
along the $H_{\rm c}$ axis has ``melted'' in comparison to the
zero-temperature results from Monte Carlo simulations, and the
distribution has broadened along the vertical $H_{\rm b}$ axis.
By comparing to Monte Carlo simulations at finite temperatures (see
below) we therefore conclude that finite temperatures and therefore
fluctuations of the magnetic moments disrupt the reversal-field memory
effect considerably even at temperatures considerably lower than the
transition temperature of the glass phase.

\section{Numerical Results}
\label{sec:numerics}

\begin{figure}[!tbp]
\includegraphics[width=\columnwidth]{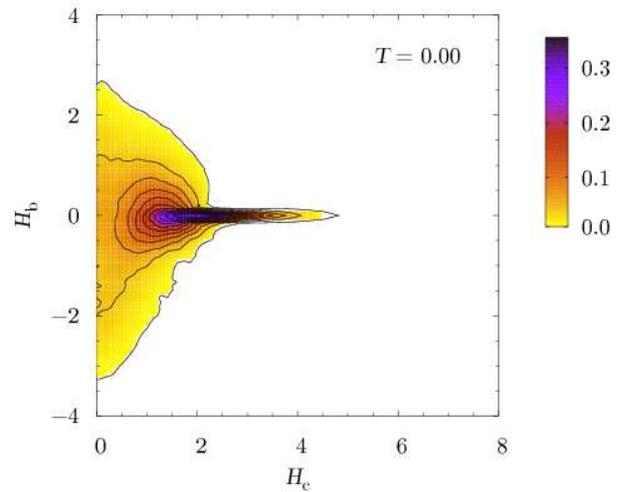}

\vspace*{-0.2cm}

\caption{(Color online)
FORC diagram of the three-dimensional Edwards-Anderson Ising spin glass
at zero temperature (see Ref.~\onlinecite{katzgraber:02b} for further details).
The data show a pronounced ridge at zero bias ($H_{\rm b} = 0$) due to
the reversal-field memory effect which is a manifestation of the spin-reversal
symmetry of the underlying Hamiltonian.
}
\label{fig:forc-T0.00}
\end{figure}

\begin{figure}[!tbp]
\includegraphics[width=\columnwidth]{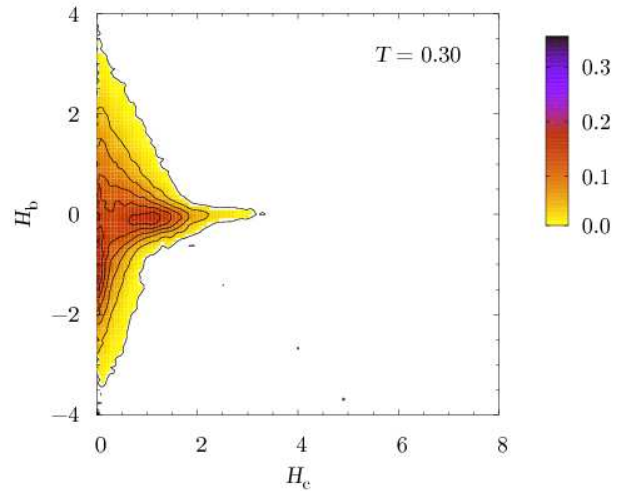}

\vspace*{-0.2cm}

\caption{(Color online)
FORC diagram of the three-dimensional Edwards-Anderson Ising spin glass
at $T = 0.30$, which corresponds to $T/T_{\rm c} \approx 0.31$. The sharp
ridge found at zero temperature (Fig.~\ref{fig:forc-T0.00}) is washed out
completely yet the asymmetry along the horizontal axis remains, in very good
agreement with the experimental data shown in Fig.~\ref{fig:forc-exp}.
}
\label{fig:forc-T0.30}
\end{figure}

The experimental results presented in Sec.~\ref{sec:experiments}
do not agree with the numerical results presented in
Ref.~\onlinecite{katzgraber:02b}. In particular, the narrow ridge
along the horizontal axis that captures the reversal-field memory
of the Edwards-Anderson Ising spin glass is washed out (see
Fig.~\ref{fig:forc-T0.00}). To better understand these findings,
we simulate the three-dimensional Edwards-Anderson Ising spin
glass\cite{edwards:75,binder:86} given by the Hamiltonian
\begin{equation}
{\mathcal H}= \sum_{\langle i,j \rangle} J_{ij}S_iS_j - H  \sum_i S_i
\label{eq:hamilton}
\end{equation}
at finite temperature. Here $S_i \in \{\pm 1\}$ represent Ising
spins on a cubic lattice of size $N = L^3$ with periodic boundary
conditions. The nearest-neighbor interactions between the spins
$J_{ij}$ are chosen from a Gaussian disorder distribution with zero
mean and standard deviation unity. $H$ represents an externally
applied field. The model has a spin-glass transition at $T_{\rm c}
\approx 0.95$.\cite{katzgraber:06}

The zero-temperature dynamics of the Edwards-Anderson Ising spin
glass is simulated by changing the external field $H$ in small steps
starting from positive saturation. After each field step we compute
the local field $h_i$ of each spin
\begin{equation}
h_i=\sum_{j} J_{ij}S_j - H \, .
\label{eq:local_field}
\end{equation}
A spin is unstable if it points opposite to its local field, i.e.,
if $h_i S_i < 0$. Randomly chosen unstable spins are flipped
and their local fields updated until all spins are
stable.\cite{comment:protocols}

At finite temperatures the system is simulated with a Monte Carlo
approach:\cite{metropolis:49,landau:97} the external field is changed
in small steps and for each field step the system is equilibrated using
heat-bath Monte Carlo.  Tests show that for $1000$ Monte Carlo lattice
sweeps at $T = 0.30$ we obtain hysteresis loops which are independent
of Monte Carlo time within error bars.  While the system is {\em not}
in full equilibrium\cite{comment:equil} neither are the experiments and
so we argue that our results describe the intrinsic nonequilibrium
nature of the finite-temperature experiments correctly.  For both zero
and finite temperatures we choose a saturation field $H_{\rm sat} =
16.0$ and perform $400$ field steps. The presented data for $20^3$
spins are averaged over $2000$ disorder realizations.

Figure \ref{fig:forc-T0.00} shows a numerical FORC diagram at
zero temperature for the Edwards-Anderson Ising spin glass (see
Ref.~\onlinecite{katzgraber:02b}). The data show a pronounced ridge
at $H_{\rm b} = 0$ which is reminiscent of the reversal-field memory
effect.  In Fig.~\ref{fig:forc-T0.30} we present data at $T = 0.30$,
a temperature (in dimensionless units) which agrees with the values
used in the experiments presented in Sec.~\ref{sec:experiments}. The
sharp horizontal ridge is completely washed out in very good
qualitative agreement with the experimental results and shows that
for the Edwards-Anderson Ising spin glass reversal-field memory is
destroyed by finite-temperature fluctuations.  This suggests that the
reversal-field memory effect\cite{katzgraber:02b} can only be probed at
temperatures much lower than can currently be achieved experimentally.

\section{Conclusions}
\label{sec:conclusions}

The hysteresis behavior of spin glasses is governed by
temperature--relaxation phenomena and field sweep rates dictate the
width and shape of the hysteresis loop at finite temperatures.  We find
fundamental differences between the field-driven hysteresis derived at
zero temperature by Monte Carlo simulations and the experimentally
as well as numerically observed behavior at finite temperature.
In particular, the FORC distributions of finite-temperature Monte
Carlo simulations and experiments on \FeMn (an Ising spin glass
material) at $T/T_{\rm c} \approx 0.31$ agree qualitatively
well (see Figs.~\ref{fig:forc-T0.30} and \ref{fig:forc-exp}).
The zero-temperature behavior which is dominated by the reversal-field
memory effect and characterized by a sharp ridge along the horizontal
axis of the FORC diagram is absent at finite temperatures.

Our results therefore show that the reversal-field  memory effect can
only be observed at temperatures close to or equal to zero which are
inaccessible experimentally with current technology. Furthermore,
the qualitative agreement between the finite-temperature Monte
Carlo data and the experimental results suggest that \FeMn is well
described by a (short-range) three-dimensional Edwards-Anderson
Ising spin glass.  It would be of interest to further characterize
other materials as well as effective models attempting to describe
these using the FORC method. This is of paramount importance for
materials for which model Hamiltonians are currently either unknown
or under debate.  For example, \FeZnF is a diluted antiferromagnet in
a field\cite{ye:02,ye:06} which is expected to be well described by
a random-field Ising model.\cite{lyuksyutov:99,sethna:93} Currently
experiments as well as simulations are being performed to characterize
this material/model using the FORC method.

\begin{acknowledgments}

We would like to thank D.~P.~Belanger, F.~Hassler, and G.~T.~Zim\'anyi
for helpful discussions.  The simulations were performed on the hreidar
cluster at ETH Z\"urich.  This work has been supported in part by
the Swiss National Science Foundation under Grant No.~PP002-114713,
the EU-RTN project DYGLAGEMEM, and the Swedish Research Council.

\end{acknowledgments}

\bibliography{comments,refs}

\begin{thebibliography}{36}
\bibitem[{com({\natexlab{b}})}]{comment:protocols}
\bibinfo{note}{An updating scheme, where the spins are sorted by their degree
  of instability (size of the local field) and then flipped in order yields the
  same results. We use random updates since these are faster to implement.}
\end{thebibliography}

\end{document}